\documentclass[sn-aps]{sn-jnl}% American Physical Society (APS) Reference Style

\jyear{2022}%

\theoremstyle{thmstyleone}%
\theoremstyle{thmstyletwo}%

\theoremstyle{thmstylethree}%

\usepackage[version=4]{mhchem} % Formula subscripts using \ce{}
\usepackage{xr}
\usepackage{soul}

\usepackage{xr}

\begin{document}{

\title[Low Thermal Conductivity Phase Change Memory Superlattices]{Low Thermal Conductivity Phase Change Memory Superlattices}

\author*[1,2]{\fnm{Jing} \sur{Ning}}\email{jing\_ning@mymail.sutd.edu.sg}
\author[3]{\fnm{Xilin} \sur{Zhou}}\email{xilinzhou@mail.sim.ac.cn}
\author[1]{\fnm{Yunzheng} \sur{Wang}}\email{yunzheng\_wang@sdu.edu.cn}
\author[4]{\fnm{Takashi} \sur{Yagi}}\email{t-yagi@aist.go.jp}
\author[5]{\fnm{Janne} \sur{Kalikka}}\email{janne.kalikka@tuni.fi}
\author[6]{\fnm{Siew Lang} \sur{Teo}}\email{sl-teo@imre.a-star.edu.sg}
\author[3]{\fnm{Zhitang} \sur{Song}}\email{ztsong@mail.sim.ac.cn}
\author[2,6]{\fnm{Michel} \sur{Bosman}}\email{msemb@nus.edu.sg}
\author*[1]{\fnm{Robert E.} \sur{Simpson}}\email{robert\_simpson@sutd.edu.sg}

\affil*[1]{\orgname{Singapore University of Technology and Design}, \orgaddress{\street{8 Somapah Road}, \postcode{487372},  \country{Singapore}}}

\affil[2]{\orgdiv{Department of Materials Science and Engineering}, \orgname{National University of Singapore}, \orgaddress{\street{9 Engineering Drive 1}, \postcode{117575}, \country{Singapore}}}

\affil[3]{\orgname{State Key Laboratory of Functional Materials for Informatics, Shanghai Institute of Microsystem and Information
Technology, Chinese Academy of Sciences}, \orgaddress{\street{865 Changning Road}, \city{Shanghai}, \postcode{200050}, \country{China}}}

\affil[4]{\orgdiv{National Metrology Institute of Japan}, \orgname{National Institute of Advanced Industrial Science and Technology}, \orgaddress{\street{Tsukuba Central 3, 1-1-1 Umezono}, \city{Tsukuba}, \state{Ibaraki}, \postcode{305-8563}, \country{Japan}}}

\affil[5]{\orgname{Computational Physics Laboratory, Tampere University}, \orgaddress{\city{Tampere}, \postcode{FI-33014}, \country{Finland}}}

\affil[6]{\orgname{Institute of Materials Research and Engineering (IMRE), Agency for Science Technology and Research (A*STAR)}, \orgaddress{\street{2 Fusionopolis Way}, \postcode{138634}, \country{Singapore}}}


\abstract{
Phase change memory devices are typically reset by melt-quenching a material to radically lower its electrical conductance.  
The high power and concomitantly high current density required to reset phase change materials is the major issue that limits the access times of 3D phase change memory architectures. 
Phase change superlattices were developed to lower the reset energy by confining the phase transition to the interface between two different phase change materials.
However, the high thermal conductivity of the superlattices means that heat is poorly confined within the phase change material, and most of the thermal energy is wasted to the surrounding materials. 
Here, we identified Ti as a useful dopant for substantially lowering the thermal conductivity of \ce{Sb2Te3-GeTe} superlattices whilst also  stabilising the layered structure from unwanted disordering.
We demonstrate via laser heating that lowering the thermal conductivity by doping the \ce{Sb2Te3} layers with Ti halves the switching energy compared to superlattices that only use interfacial phase change transitions and strain engineering. 
The thermally optimized superlattice has $(0~0~l)$ crystallographic orientation yet a thermal conductivity of just 0.25~W$\cdot$m$^{-1}\cdot$K$^{-1}$ in the "on" (set) state. 
Prototype phase change memory devices that incorporate this Ti-doped superlattice switch faster and and at a substantially lower voltage than the undoped superlattice.
During switching the Ti-doped \ce{Sb2Te3} layers remain stable within the superlattice and  only the Ge atoms are active and undergo interfacial phase transitions.
In conclusion, we show the potential of thermally optimised \ce{Sb2Te3-GeTe} superlattices for a new generation of energy-efficient electrical and optical phase change memory.}




\maketitle
\pagebreak

\section{Introduction}

Data is being created, stored, and processed at a staggering rate. 
Indeed, the world's data production doubles in size every two years and by 2025 it is predicted to reach a whopping 181 zettabytes per year~\citep{statista}, whilst the data storage demand is predicted to be 30.23 zettabytes in 2030~\citep{koot2021usage}.
This exponential growth in data production and storage is fueling innovations in materials, device scaling, and novel data storage concepts. 
Consequently, the amount of energy consumed to store the world's data is expected to have a 17\% reduction from 18.33 TWh in 2016 to 15.23 TWh in 2030~\citep{koot2021usage}.

High speed and low energy prototype resistive~\citep{chen2020reram}, ferroelectric~\citep{kim2019ferroelectric}, magnetoelectric~\citep{amiri2015electric}, and phase change~\citep{zhang2019designing} memories are all currently being developed.
Phase change memories are currently leading the way and are being produced commercially.
Phase change memory encodes data in the local structural order of a phase change material (PCM), and the data is read as a substantial electrical resistance change, which results from the different structural states.
Typically, the amorphous and crystalline states of the PCM are used to encode binary information into the material.
The amorphous to crystalline (SET) transition is induced by heating the material for a duration sufficiently long that the atoms can overcome an energy barrier and move into a lower energy configuration.
The reversible transition from the crystalline (SET) phase to the amorphous (RESET) phase is realized by melt-quenching the material at a rate higher than the crystallization rate, such that the atoms are frozen in a higher energy state.
Many chalcogenide materials, especially those along the \ce{Sb2Te3-GeTe} pseudo-binary tie-line in the Ge--Sb--Te ternary diagram exhibit rapid recrystallization that results in a substantial electrical and optical contrast. 
Arguably, \ce{Ge2Sb2Te5} is the most famous of these materials. 
\ce{Ge2Sb2Te5}-based compounds have been widely studied and commercially applied in DVD-RAM optical discs~\citep{ohta2001phase} and the 3D X'point technology electrical memory~\citep{joelhruska}.

Despite their commercial success, the main drawback with PCM-based electrical data storage is the high RESET power.
Generally, the PCM RESET current needs to be lower than the ON-state current of the memory cell selector.
To achieve high data densities, two terminal ovonic threshold selectors (OTS) are replacing their three terminal transistor counterparts~\citep{zhang2015one}. 
However, the current density of an OTS tends to be lower than transistor selectors, and most selectors can only pass a current up to $\sim$10~MA/cm$^{2}$~\citep{zhu2019ovonic}.
Therefore, the RESET current density of the PCM must be limited below the ON-state OTS current, which is a big problem because longer pulse times are required to reset the PCM as a compromise, and this increases the phase change memory access time.
We note, however, that most of the heat generated during the RESET process is dissipated into the surroundings and only 1\% of the energy is actually used to drive the PCM phase transitions~\citep{sadeghipour2006phase}.
By decreasing the heat dissipated from the PCM element, there is a clear opportunity to decrease the RESET current and concomitantly decrease the PCM device access time.  

Substantial research has focused on decreasing the energy consumed to switch PCMs, and three approaches have emerged to improve the switching efficiency: 
(1) Decrease the thermal conductivity; 
A variety of dopants to lower the switching current have been investigated, such as C~\citep{zhou2014understanding}, Cr~\citep{wang2015cr}, Sc~\citep{wang2018scandium}, and Ti~\citep{zhu2014uniform}. 
These dopant atoms diffuse into the PCMs to form local defects or distortions, which results in low thermal conductivity.
The generated heat is confined in the PCM, which in turn leads to low energy switching. 
Similarly, a "superlattice-like" structure, which consists of non-epitaxial \ce{Sb2Te3} and GeTe stacked layers, was proposed to lower the switching energy by decreasing the thermal conductivity via interfacial phonon scattering~\citep{chong2006phase}. 
More recently, several reports have demonstrated how two-dimensional crystals, such as graphene, \ce{MoS2}, and \ce{WS2} can be placed at the interface between the PCM and the electrode or substrate to confine heat through increased thermal boundary resistance\citep{Ahn15, Neumann19, ning2022}.
(2) Decrease entropic losses; 
Interfacial phase change materials (iPCMs) improve the switching performance by limiting the movement of the atoms during the phase transition. 
Less movement freedom reduces entropic losses~\citep{simpson2011interfacial} and concomitantly decreases the energy required to switch the material. 
%Typically, atoms can move in all three dimensions, but the layered iPCM superlattice structure confines their movement to one or two dimensions, which significantly reduces entropic losses~\citep{simpson2011interfacial} and concomitantly decreases the energy required to switch the material. 
The iPCM structure, which also has a \ce{Sb2Te3-GeTe} layered structure but with a high degree of crystallographic orientation and lattice matching, was reported to have a higher thermal conductivity than both the \ce{Ge2Sb2Te5} and the superlattice-like structure. 
Recently, however, the cross-plane thermal conductivity of the \ce{Sb2Te3-GeTe} superlattices was shown to decease with increasing number of \ce{Sb2Te3-GeTe} interfaces~\citep{kwon2021uncovering}.  
We conclude, therefore, that both thermal confinement effects and smaller entropic losses contribute to the lower switching energy seen in iPCM devices. 
%It also indicates that if an iPCM structure can be developed that combines reduced entropic losses with low thermal conductivity, then further switching energy enhancements will be possible.
(3) Decrease the activation energy for the Ge atomic transitions; 
that the activation energy for atomic transitions in the iPCM structure can be controlled using strain engineering; indeed, straining the GeTe layer by just 2.1\% leads to substantial decrease in the switching voltage.
The Ge atom interfacial transition is an activated process. 
Biaxially straining the superlattice makes the Ge atom transition easier, thus lowering the switching energy of iPCM devices~\citep{kalikka2016strain,zhou2017avalanche}.
All these methods can improve the switching performance to some extent, however, until now they have not been combined into a thermally-engineered, strain-engineered, interface-engineered "Super iPCM".

In this work, a highly efficient [Ti-\ce{Sb2Te3}]$_{\rm{4nm}}$--[\ce{GeTe}]$_{\rm{1nm}}$ iPCM superlattice structure, which exhibits interfacial atomic transitions, strain-engineered layers, and low thermal conductivity is designed and demonstrated in a prototype memory cell. 
Adding Ti into bulk \ce{Sb2Te3} primarily increases phonon scattering from point defects and consequently decreases the thermal conductivity~\citep{dravsar2005transport}.
If this same effect exists in \ce{Sb2Te3} textured thin films, then one should expect a substantial decrease in the iPCM thermal conductivity.
This decrease should be especially dramatic in the strain engineered iPCM structure where the \ce{Sb2Te3} layers are four times thicker than the GeTe layers.
It is therefore hypothesized that doping Ti into Sb sites in the \ce{Sb2Te3} crystal will be an effective way to lower the thermal conductivity of the whole superlattice structure.
Clearly, the Ti atoms must not interfere with the Ge atomic interfacial transitions, nor with the strain in the GeTe layers for the structure to maintain lower entropic losses and lower activation energy.

Herein, we show radically lower switching energies are possible by doping the \ce{Sb2Te3} layers with Ti.
The Ti-doped \ce{Sb2Te3} superlattice structure can retain a highly out-of-plane crystallographic orientation for Ti concentrations up to 3.6 at.\%.
The thermal conductivity of the \ce{Sb2Te3-GeTe} layers is lowered by over 50\% when 3.6 at.\% Ti is doped into the structure.
Density Functional Molecular Dynamics (DF/MD) simulations confirm that the GeTe layers disorder at a lower temperature than the Ti-\ce{Sb2Te3} layers, which is a fingerprint of the interfacial switching and demonstrates the stability of the Ti-\ce{Sb2Te3} layers~\citep{kalikka2016strain}.
Subsequently, we study the switching performance of this super iPCM using laser pulses.
The Ti-doped iPCM superlattice structure RESET switching energy is halved, and the pulse widths of 10 ns are enough to switch the material.
When this Ti-doped iPCM is incorporated into prototype memory cells, we show that the programming voltage is reduced by 30\%.
Moreover, the iPCM device for multilevel data storage application is demonstrated showing well separated four resistance states by doping \ce{TiTe2} into the \ce{Sb2Te3} scaffold material of the iPCM.
These results show that carefully designing the iPCM thermal properties is a viable method to meet the switching speed and energy requirements of future high speed phase change memory architectures, such as those using OTS selectors.

\section{Results and Discussion}
\subsection{Highly \textbf{(0 0 l)} orientation}
\label{structure}
The efficient switching in iPCM relies on the interfaces between the \ce{Sb2Te3} and \ce{GeTe} blocks being terminated with Te, which enables van der Waals (vdW) bonding between the blocks, as shown in Fig.~\ref{fig_1}(a). 
It is, therefore, critical to ensure that this feature of the superlattice structure is maintained when any dopants are added to one of the layers.
 
A series of Ti-doped \ce{Sb2Te3} were grown to study how the Ti dopant concentration influences the vdW layered structure.
Low-power RF-co-sputtering from \ce{TiTe2} and \ce{Sb2Te3} targets was used to grow a series of Ti$_x$--(\ce{Sb2Te3})$_{1-x}$ films at 300~$^{\circ}$C. 
These Ti$_x$--(\ce{Sb2Te3})$_{1-x}$ films maintained a highly $(0~0~l)$ orientated structure for all doping concentrations studied.
However, Ti-doped \ce{Sb2Te3} without phase separation was only possible within a narrow doping Ti concentration window, as shown by their XRD patterns in supplementary Fig.~\ref{fig_S_XRD_Raman}(a). 
This is due to there being the following two chemical equilibria in \ce{Sb2Te3} and \ce{TiTe2} co-sputtering:
\begin{equation}
	\ce{\frac{5}{2} Sb2Te3 + TiTe2 <=>T[\Delta] \frac{1}{2}Te + TiSb5Te9}
	\label{R1}
\end{equation}
\begin{equation}
	\ce{\frac{1}{10} Te + TiSb5Te9 <=>T[\Delta] \frac{5}{2} Sb2Te3 + \frac{1}{5}Ti5Te8}
	\label{R2}
\end{equation}
At relatively low Ti concentrations, free energy and formation energy analysis reveals that the \ce{TiTe2} enters vdW gap and then Ti atoms substitute Sb atoms to form \ce{TiSb5Te9}, see supplementary Fig.~\ref{fig_S_energy}. 
This Ti substitution for Sb picture is consistent with the experimentally observed decrease in hole concentration due to Ti doping \ce{Sb2Te3} single crystals~\citep{dravsar2005transport}.
Crystal Orbital Hamiltonian Population (COHP) analysis reveals that this Ti for Sb substitution increases the stability of the superlattice, see supplementary Fig.~\ref{fig_S_COHP}.
Again, an observation that is consistent with Ti dopants releasing an electrons into the conduction band that can recombine with holes~\citep{dravsar2005transport} to decrease the overall energy of the system and increase the stability.  
As the concentration of \ce{TiTe2} increases, as per Reaction~\ref{R1}, the structure phase separates into \ce{TiSb5Te9}, and free \ce{Te}.
Conveniently, however, Te has a low sticking coefficient at elevated temperatures and this facilitates the synthesis of \ce{TiSb5Te9} textured films, as per Reaction~\ref{R2}.
Naturally, diffraction peaks associated with the \ce{Te} or \ce{Ti5Te8} are seen when \ce{TiTe2} is sputtered at the highest rates because separated \ce{Te} or \ce{Ti5Te8} phase exists in the film.
The texture of Ti$_x$--(\ce{Sb2Te3})$_{1-x}$ retains a layered $(0~0~l)$ preferred orientation for 1.7 $\le x \le$3.6~at.\% Ti, and 
peaks associated with phase separation into \ce{TiTe2}, Te and \ce{Ti5Te8} are not present in the XRD patterns.
In section \ref{sec:therm}, we see that $x=$~3.6~at\% Ti has the lowest thermal conductivity. 
This is important because it shows that low thermal conductivity vdW layered structures are possible without phase separation.

We grew high quality Ti$_{3.6}$--(\ce{Sb2Te3})$_{96.4}$--GeTe superlatitcse, which were hypothesized to have a lower thermal conductivity than its undoped \ce{Sb2Te3-GeTe} counterpart. 
The hexagonal crystal structure has $(0~0~l)$ texture with alternating layers of Ti$_{3.6}$--(\ce{Sb2Te3})$_{96.4}$ and GeTe stacked along the [1~1~1] direction of their rhombohedral primitive cells. 
Compared with the XRD patterns of the \ce{Ti-Sb2Te3} layers, the superlattice has two extra diffraction peaks at 25.1$^{\circ}$ and 51.5$^{\circ}$ that correspond to the $(0~0~l)$ reflections of hexagonal GeTe.
Both doped and undoped superlattice films retained a high $(0~0~l)$ crystallographic orientation, as shown in Fig.~\ref{fig_1}(b).
The slightly lower intensity and broader diffraction peaks in the Ti-doped superlattice structure is due to a smaller in-plane grain size and local distortions induced by Ti dopants.
The XRD patterns clearly show that the Ti-doped superlattice film still retains the preferred out-of-plane orientation.

To confirm the influence of Ti on the local crystal structure, Raman spectra were measured with and without the Ti dopant.
The spectra are shown in Fig.~\ref{fig_1}(c).
Here, we highlight the \ce{Sb2Te3} and GeTe phonon modes in black and red dot-and-dash lines respectively. 
$A^1_{1g}$(69 cm$^{-1}$), $E^2_{g}$(109 cm$^{-1}$), and $E^2_{1g}$(167 cm$^{-1}$) modes seen in the \ce{Sb2Te3} film have been reported before~\citep{zhou2017avalanche}.
The Ti-doped \ce{Sb2Te3} layers retain the characteristic peaks of \ce{Sb2Te3} with a weak shoulder at 120~cm$^{-1}$.
This signal is from in-plane vibrations of Ti--Te or Te--Te bonds. 
Hence, we conclude that the Ti atoms successfully substitute Sb atoms within the \ce{Sb2Te3} quintuple blocks.
Note, the evolution of the \ce{Sb2Te3} vibrational modes with Ti concentration is shown in supplementary Fig.~\ref{fig_S_XRD_Raman}(b). 
New modes between 120~cm$^{-1}$ and 130~cm$^{-1}$ emerge in the \ce{Sb2Te3} Raman spectra when \ce{TiTe2} is co-sputtered with \ce{Sb2Te3}.  
The corresponding XRD patterns in supplementary Fig.~\ref{fig_S_XRD_Raman}(a) exhibit a similar diffraction peak that appears and disappears according to phase separation into \ce{TiTe2}, \ce{Te}, and \ce{Ti5Te8} as the concentration of \ce{TiTe2} increases.
The characteristic peaks associated with layered \ce{Sb2Te3} are well-preserved when the Ti concentration is in the range of 1.7--3.6~at.\% and phase separation does not occur in this doping range.
%Supplementary Fig S.X shows that increasing Te concentration beyond 3.6\% causes phase separation into Te or \ce{Ti5Te8}.
%For lower concentrations, the characteristic peaks associated with layered \ce{Sb2Te3} are well-preserved.

Importantly, the Raman spectra also indicate that Ti helps to prevent intermixing of the superlattice layers.
Comparing the Ti$_{3.6}$--(\ce{Sb2Te3})$_{96.4}$--GeTe and the undoped \ce{Sb2Te3-GeTe} superlattice structures, we see that the E-mode at 79 cm$^{-1}$ from GeTe is clearly visible in the Ti$_{3.6}$--(\ce{Sb2Te3})$_{96.4}$--GeTe superlattice structure and less prominent in the undoped structure. 
The E-mode for GeTe is an in-plane vibration and its presence indicates that its in-plane structure has not been disturbed.  
Preventing intermixing is important since the low energy switching in iPCM is based on confining the Ge-atoms movement at the interfaces rather than diffusing through the whole structure. 
Thus, the loss of the layered structure increases the RESET switching current of PCM devices~\citep{tominaga2020intermixing,simpson2011interfacial,khan2022unveiling}.

The layered structure of the Ti$_{3.6}$--(\ce{Sb2Te3})$_{96.4}$--GeTe superlattice is also shown in the cross-sectional STEM image.
The superlattice layers clearly form blocks that are separated by vdW gaps, as shown in Fig.~\ref{fig_1}(d).
The corresponding FFT image shows a series of streaks consisting of more closely spaced spots. 
This is a typical pattern for a large crystal unit cell exhibiting periodic superlattice layers in real space.
The electron diffraction FFT analysis reveals out-of-plane layering with a period of 46.37$\pm$0.20~\AA, which agrees well with the simulated c-axis of the corresponding unit cell (46.65~\AA), as seen in Fig~\ref{fig_1}(a). %(46.65~\AA in simulated structure)
We conclude that the fabricated Ti$_{3.6}$--(\ce{Sb2Te3})$_{96.4}$--GeTe superlattice structure exhibits $(0~0~l)$ layering and is similar to the model presented in Fig~\ref{fig_1}(a), which we used for our DFT calculations.

We have shown unequivocally using XRD, Raman spectroscopy, and STEM imaging that the Ti$_{3.6}$--(\ce{Sb2Te3})$_{96.4}$--GeTe superlattice exhibits a high quality layered structure.
The fact that Ti dopants can be introduced into the superlattice quintuple \ce{Sb2Te3} layer opens new possibilities for tuning the superlattice properties. 
In the subsequent section, we will show how the thermal conductivity of the superlattice is substantially lower than its undoped counterpart.

\subsection{Lowering the thermal conductivity} \label{sec:therm}
PCMs with a low thermal conductivity improve the switching energy efficiency by confining the generated Joule heat within the PCM rather than allowing the heat to diffuse into the surrounding materials. 
Therefore, we deliberately doped the \ce{Sb2Te3} layers of the superlattice structure with Ti to improve its switching energy performance.  
Ti was chosen because it was found to decrease the thermal conductivity of \ce{Sb2Te3} bulk single crystals.
Herein, we report time-dependent thermal reflectance (TDTR) measurements that show how Ti has a dramatic effect on the thermal conductivity of $(0~0~l)$ textured \ce{Sb2Te3} thin films.
The thermal conductivity of \ce{Sb2Te3} exhibits a parabolic dependence on the Ti concentration, as shown in Fig.~\ref{fig_2}(a).
The measured thermal conductivity of \ce{Sb2Te3} was $\sim$0.64~W$\cdot$m$^{-1}\cdot$K$^{-1}$, which is consistent with previously reported values in the literature~\citep{rowe2018crc,lensch2012electrodeposition,shen2017enhancing}.
The lowest thermal conductivity was observed for Ti concentrations between 3.6~at\% and 6~at\%.
In section \ref{structure}, we found that for Ti concentrations greater than 3.6 at.\% resulted in phase separation.
Therefore, the 3.6~at.\% Ti-doped sample exhibits both a textured structure and low thermal conductivity, which when incorporated into a superlattice structure should allow both interfacial atomic transitions and efficient heat trapping.

The effect of Ti on the superlattice thermal conductivity was confirmed by comparing the thermal conductivity of undoped \ce{Sb2Te3-GeTe} and Ti$_{3.6}$--(\ce{Sb2Te3})$_{96.4}$--GeTe superlattice structures. 
The thermal conductivity of the undoped \ce{Sb2Te3-GeTe} superlattice structure was measured to be 0.49~W$\cdot$m$^{-1}\cdot$K$^{-1}$ in the set (on) state whilst the Ti$_{3.6}$--(\ce{Sb2Te3})$_{96.4}$--GeTe superlattice thermal conductivity was more than halved, to just 0.24~W$\cdot$m$^{-1}\cdot$K$^{-1}$, as shown in Fig.~\ref{fig_2}(a).
The decreasing thermal conductivity is generated from the Ti-induced point defects and local distortions within its crystalline grains (see Fig.~\ref{fig_1}(d))~\citep{dravsar2005transport,zhu2014uniform}.
%Moreover, for Ti-doped and undoped samples with similar thickness, this effect is compounded by less intermixing due to the increased the number of vdW gaps, resulting in a lower thermal conductivity. \hl{I removed this sentance because Eric Pop's group showed results indicating the the vdW gaps do not signifficantly effect the thermal conductivity).}
These results suggest that the Ti doped superlattice should switch substantially more efficiently than its undoped counterpart.

The topography of the Ti-doped superlattice film suggests the presence of much smaller crystal grain sizes than that for the undoped superlattice. %and the additional grain boundaries lower the thermal conductivity.
The mean grain diameter of the doped superlattice structure is $\sim$20--40~nm whilst it is $\sim$80--110~nm in the undoped superlattice, as shown in Fig.~\ref{fig_2}(b) and (c). 
Similar features are observed in the SEM images, which are included in the supplementary section for the interested reader, see Fig.~\ref{fig_S_SEM}.
The smaller grains improve the film uniformity and result in smoother films. 
Indeed, the AFM-measured RMS roughness decreased from 3.62~nm for the undoped structure to 1.47~nm for the doped structure. 
Smoother films are preferred when building multilayered 3D device architectures because the effect of layer roughness needs to be minimized.  

\subsection{Low Switching Energy Ti-doped superlattices}
An important design consideration for iPCM superalttice structures is an active layer that switches and is embedded in a stable scaffold~\citep{kalikka2016strain}.  
Indeed, the GeTe layers embedded within the \ce{Sb2Te3} or \ce{Sb2Te1} scaffold exhibit premelting, which is similar to amorphization but limited to the GeTe layers. 
The \ce{Sb2Te3} films remain crystalline and the layered structure is uncompromised~\citep{kalikka2016strain,zhou2016phase}.
It is, therefore, important to confirm whether the Ti$_{3.6}$--(\ce{Sb2Te3})$_{96.4}$--GeTe superlattice also exhibits premelting.

The melting behavior of the Ti$_{3.6}$--(\ce{Sb2Te3})$_{96.4}$--GeTe superlattice was studied using DF/MD simulations from 450~K to 1200~K in 375~ps, as shown in Fig.~\ref{fig_3}. 
The crystallinity of the individual GeTe and \ce{Ti-Sb2Te3} layers was calculated as a function of temperature and time using the Steinhardt bond-orientation parameter~\citep{steinhardt1983bond}. 
The initial structure was fully ordered and the crystallinity was equal to 1, whilst it fully disorders when the temperature exceeds 1135~K, as shown in the inset of Fig.~\ref{fig_3}.
We also see that the GeTe layer disorders at a much lower temperature than the Ti$_{3.6}$--(\ce{Sb2Te3})$_{96.4}$ layer; i.e. the superlattice does indeed exhibit GeTe premelting.
The GeTe starts disordering at 650~K, whilst the Ti$_{3.6}$--(\ce{Sb2Te3})$_{96.4}$ scaffold layers remain highly ordered up to $\sim$850~K.
The crystallinity of GeTe is reduced by 20\% to 0.8 at 1025~K whilst the Ti$_{3.6}$--(\ce{Sb2Te3})$_{96.4}$ layer cannot reach this level until the whole structure completely melted.
% only reached this level of disorder at $\sim$1200~K in fitted curve, where the whole structure completely melted.
This result provides strong evidence that Ti-doped \ce{Sb2Te3} can efficiently switch by premelting the GeTe layers within the superlattice. 
Importantly, the Ti$_{3.6}$--(\ce{Sb2Te3})$_{96.4}$--GeTe superlattice exhibits a substantially lower thermal conductivity than the undoped superlattice yet it maintains the GeTe premelting characteristic, both of which will enable low-energy switching.
 
A PCM that can switch at high speed with low energy pulses is important for short access times in memory devices.
We used a pump-probe laser system to measure the switching energy and time of the doped and undoped superlattices~\citep{behera2017laser}.
The as-grown crystalline Ti-doped \ce{Sb2Te3-GeTe} superlattice films were irradiated with  laser pulses with different durations and powers.
The optical reflectivity of the film decreased as the laser energy deposited into the film increased.
This is indicating a phase transition between the two structural phases. 
The laser pulse power-time-reflectivity for the \ce{Sb2Te3}-GeTe and Ti$_{3.6}$--(\ce{Sb2Te3})$_{96.4}$--GeTe superlattices are presented in Fig.~\ref{fig_4}(a) and (b) respectively.

The reset switching energy of the Ti$_{3.6}$--(\ce{Sb2Te3})$_{96.4}$--GeTe superlattice is nearly three times lower than that of the undoped \ce{Sb2Te3}-GeTe superlattice.
The laser power-time-reflectivity curves, which are shown in Fig.~\ref{fig_4}, reveal that the undoped superlattice required $\sim$28.72~mW, 60~ns ($1.72~\mu J$) RESET pulses. 
In contrast, the Ti$_{3.6}$--(\ce{Sb2Te3})$_{96.4}$--GeTe superlattice was reset with just $\sim$13.45~mW and 50~ns ($0.67~\mu J$) pulses; which corresponds to a 2.6$\times$ reduction in switching energy. 
The only difference between the two superlattices is their thermal conductivity due to  Ti doping the low switching energy structure.
Fig.~\ref{fig_2} shows that the thermal conductivity of the Ti$_{3.6}$--(\ce{Sb2Te3})$_{96.4}$--GeTe superlattice is approximately half that of the undoped superlattice.

To a first approximation, the characteristic time for heat conduction through a material is $\Delta t \approx \frac{x^2 \rho C_p}{\kappa}$, where $x$ is the distance from the laser heat source to the substrate, $\kappa$ is the thermal conductivity, $\rho$ is the density and $C_p$ is the heat capacity.
Thus, the distance over which heat diffuses depends on the square root of the thermal conductivity. 
Here, the thermal conductivity of the doped superlattice is half that of the undoped superlattice. 
Therefore, the Ti$_{3.6}$--(\ce{Sb2Te3})$_{96.4}$--GeTe superlattice confines the heat within a diameter that is approximately 1.4 smaller than that of the undoped superlattice.
Since heating is governed by an imbalance in the heat flux in and out of the film, reducing the rate that heat can diffuse out of the structure leads to heat confinement and enables the GeTe layers to premelt-disorder with lower input energies.
%\hl{This discussion may not be reasonable enough. The claim "all else being equal" is not true since the density and heat capacity of the film can not be the same after Ti doping.}

The Ti$_{3.6}$--(\ce{Sb2Te3})$_{96.4}$--GeTe superlattice layered structure is stable even after laser resetting.
The superlattice is remarkably stable and even after resetting at the $\sim$33.42~mW power limit of our laser, subsequent cross-sectional TEM analysis shows a layered structure, see Fig.~\ref{fig_4}(c).
The streaks and additional spots in the TEM selected area electron diffraction pattern (inset) also show that the superlattice structure remains layered and is similar to the as-grown superlattice. 
The uncapped structure could be cycled between the SET and RESET states more than 4000 times using 13.45~mW for 50~ns RESET pulses and 9.42~mW for 300~ns SET pulses. 
These results are provided in Fig~\ref{fig_S_cycle} of the supplementary section. 
We expect higher cycle abilities in electrical memory devices because the structure is protected from the atmosphere by the surrounding materials.

The Ti$_{3.6}$--(\ce{Sb2Te3})$_{96.4}$--GeTe superlattice is heated more rapidly and reaches a substantially higher equilibrium temperature than the undoped \ce{Sb2Te3-GeTe} superlattice when irradiated with the same laser pulse parameter.
The minimum laser powers needed to RESET the superlattices with 70 ns pulses were found from the power-time-reflectivity matrices, see Fig.~\ref{fig_4}(a) and (b).
They were respectively 28.72~mW and 13.45~mW for the \ce{Sb2Te3-GeTe} and Ti$_{3.6}$--(\ce{Sb2Te3})$_{96.4}$--GeTe superlattices.
The finite-difference method  was then used to simulate the cross-sectional temperature distribution in the superlattices as a function of time during these RESET pulses. 
Both superlattice samples reached a similar temperature (over 900~K) after 70~ns, as shown in Fig.~\ref{fig_S_heat}, which is sufficient for premelt disordering, hence we observe a transition to the RESET reflectivity state in Fig.~\ref{fig_4}(a) and (b).
Importantly, however, the laser power used to RESET the Ti$_{3.6}$--(\ce{Sb2Te3})$_{96.4}$--GeTe superlattice was only 47\% of that needed to RESET the undoped structure; a substantial switching energy efficiency improvement.
To emphasize the difference in heating efficiency, the temperature rise in the two structures due to irradiating with the same 70~ns for 13.45~mW laser pulses was simulated.
The equilibrium temperature difference between the two superlattices was 321~K, as shown in Fig.~\ref{fig_4}(d).
The temperature distributions at 70~ns are also inset into Fig.~\ref{fig_4}(d).
The undoped superlattice temperature does not even reach 600~K and remains well below the GeTe premelt disordering temperature.
Therefore, the undoped superlattice cannot RESET with 70~ns 13.45~mW pulses. 
These simulations highlight the substantial difference in heating efficiency and agree well with our laser switching experiments.

The substantial improvement in heating efficiency implies a reduction in the switching voltage when these superlattices are employed in electrical phase change memory cells.
The reversible resistance-voltage (R-V) switching measurement for the undoped \ce{Sb2Te3-GeTe} and Ti$_{3.6}$--(\ce{Sb2Te3})$_{96.4}$--GeTe superlattices devices is given in Fig.~\ref{fig_4}(e) and (f).
The pulse duration on the R-V switching curves was 20~ns and 40~ns.
The Ti$_{3.6}$--(\ce{Sb2Te3})$_{96.4}$--GeTe superlattices device showed over two orders of magnitude change in electrical resistance between the SET and RESET states, which is larger than that of undoped superlattice device.
The Ti-doped iPCM cell started to switch with 20~ns pulses and two orders of magnitude change was possible when the pulse time was 40~ns. 
With 40~ns pulses, the undoped superlattice (1.6$\pm$0.1~V) and Ti$_{3.6}$--(\ce{Sb2Te3})$_{96.4}$--GeTe superlattice  can be set with a similar programming voltage of 1.65$\pm$0.15~V.
The thermal effect has an even greater influence on the RESET voltage because of the higher temperature needed for premelt-disordering.
The undoped superlattice resets at 3.4$\pm$0.1~V whereas the Ti$_{3.6}$--(\ce{Sb2Te3})$_{96.4}$--GeTe structure resets at 2.65$\pm$0.25~V with 40~ns pulses.
The less abrupt switching occurred in Ti$_{3.6}$--(\ce{Sb2Te3})$_{96.4}$--GeTe superlattices device, showing the possibility for multi-level operation. 
In the supplementary section, we have also included the DC current-voltage (I-V) curves, see Fig.~\ref{fig_S_IV_curve}.
The superlattices exhibit ovonic threshold switching~\citep{ovshinsky1968reversible} whereby the disordered GeTe layers in the RESET state produce a voltage snapback and increase in conductivity at a threshold voltage V$_{th}$. 
The V$_{th}$ of the Ti$_{3.6}$--(\ce{Sb2Te3})$_{96.4}$--GeTe superlattice devices was lower than that of the undoped devices. 
This lower OTS voltage is indicative of a structure with more charged defects, which is consistent with Ti atoms substituting Sb atoms to producing positively charged defects~\citep{dravsar2005transport}. 
All of these electrical switching results show that Ti improves the heating efficiency of the superlattice without negatively affecting the electrical resistance change.

%\hl{[[I suggest we did not complicate the story with this result. Publish another paper!]}The Ti-doped \ce{Sb2Te3-GeTe} superlattice exhibit multi-level switching with the resistance levels depending on pulse voltage and duration.

%\section{Discussion}
%
%Discussions should be brief and focused. In some disciplines use of Discussion or `Conclusion' is interchangeable. It is not mandatory to use both. Some journals prefer a section `Results and Discussion' followed by a section `Conclusion'. Please refer to Journal-level guidance for any specific requirements. 

\section{Conclusion}

%The iPCM superlattice is able to switch with an order of magnitude lower energy than its alloy counterpart by reducing entropic losses.
We have shown that the switching energy of interfacial phase change memory superlattices can be further improved by doping the \ce{Sb2Te3} layers with elements that lower their thermal conductivity.
Indeed, when the thermal conductivity of the superlattice is halved, the RESET switching energy also decreases by 50\%.
As with the undoped superlattice, the GeTe layers in the Ti-doped superlattice premelt-disorder before the superlattice completely amorphizes. 
Indeed, the layered structure is robust and present even after switching into the high resistance state.

Thermal conductivity is an important knob that can be used to enhance the iPCM switching energy efficiency.
Titanium is just one dopant that we have identified that can lower the superlattice thermal conductivity without influencing the crystal structure and the interfacial switching mechanism.
We suspect that other elements can also provide similar benefits. 
Generally, this work highlights the importance of co-optimizing the crystal structure, thermal conductivity, entropy, and strain to achieve a low energy switching phase change material. 
%A more than 300~K difference in temperature rise in optical switching when Ti in the \ce{Sb2Te3} layer.
We expect that these super-iPCM materials will be applied in the next generation of high speed, energy efficient nonvolatile memory technologies.

% Figures/tables and captions
% Permission statements are required for all figures reproduced or adapted from previously published articles/sources. Please also ensure that all necessary permissions to reproduce images have been received
% Please remove these statements for original figures

\begin{figure}[!htbp]
  \includegraphics[width=\textwidth]{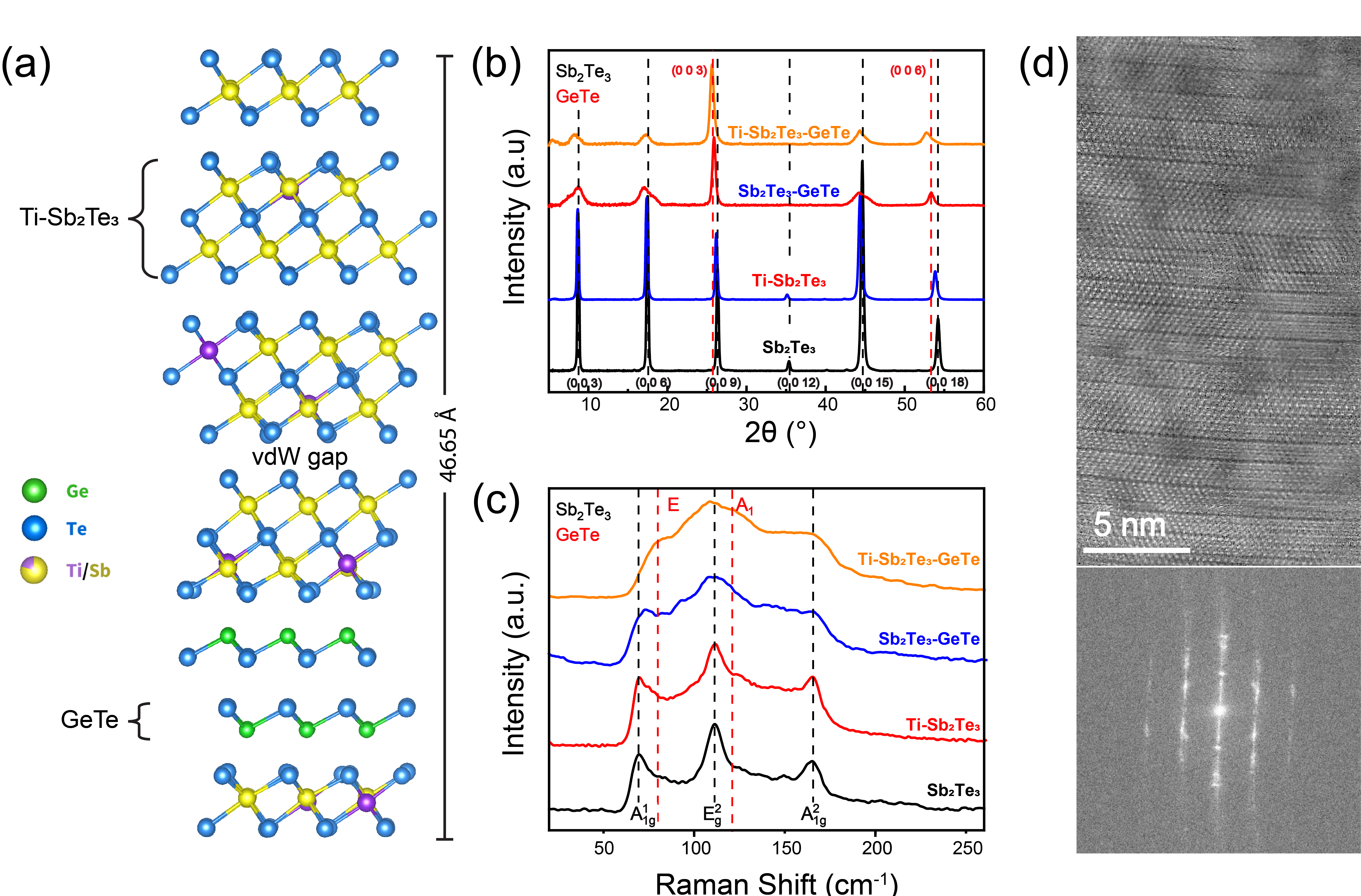}
  \caption{\textbf{Structure of the Ti-doped superlattice.}
  (a) Schematic of the [Ti-\ce{Sb2Te3}]$_{4nm}$-[\ce{GeTe}]$_{1nm}$ superlattice crystal structure. 
  (b) XRD patterns of pure \ce{Sb2Te3} and Ti-doped \ce{Sb2Te3} and their superlattice films. 
  (c) Raman spectra of pure \ce{Sb2Te3} and Ti-doped \ce{Sb2Te3} and their superlattice films.
  (d) Cross-section TEM image of [Ti-\ce{Sb2Te3}]-[\ce{GeTe}] superlattice layers and the corresponding FFT image of this area.
}
  \label{fig_1}
\end{figure}

\begin{figure}[!htbp]
  \includegraphics[width=\linewidth]{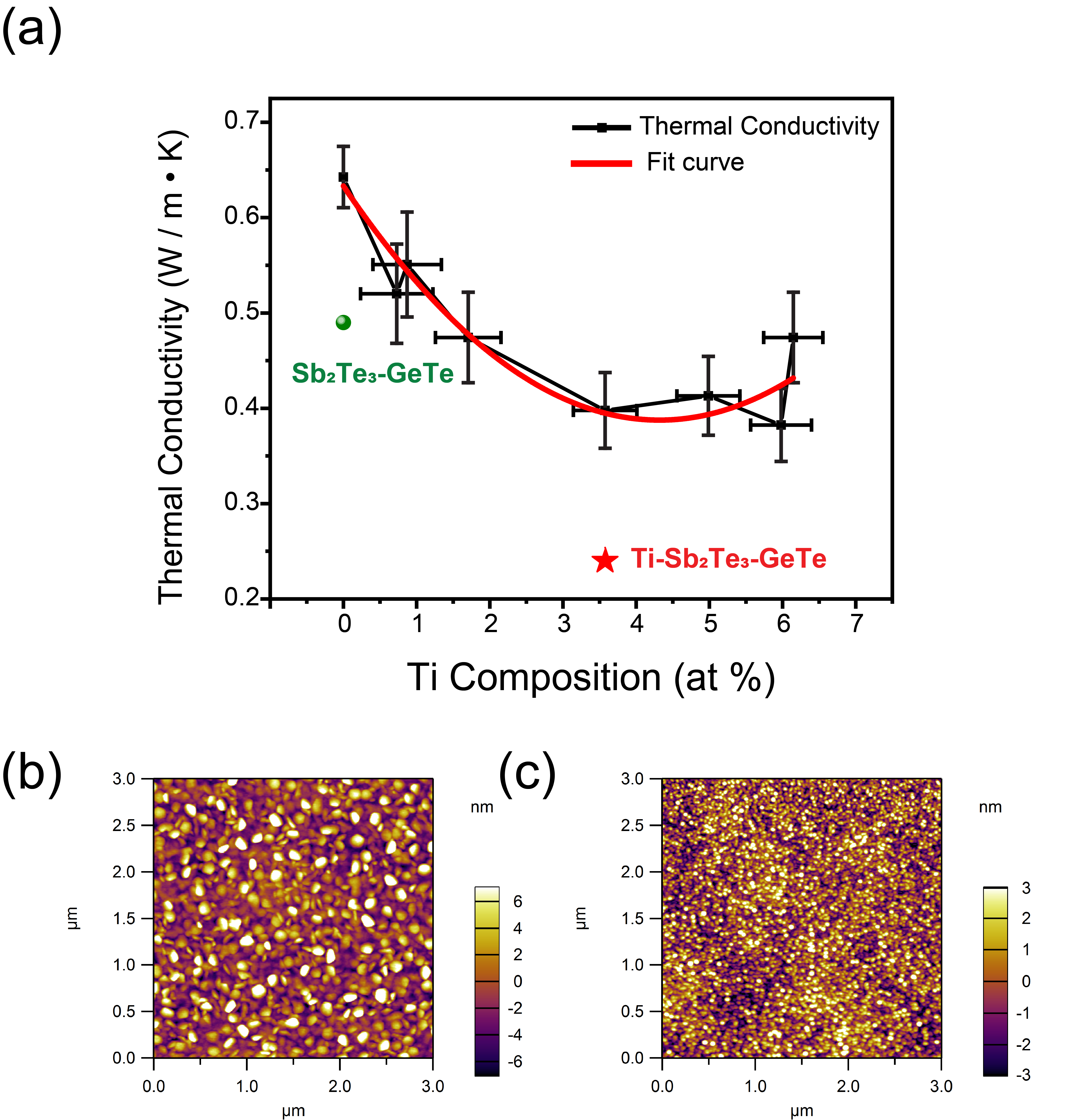}
  \caption{
  (a) Thermal conductivity of Ti-doped \ce{Sb2Te3} films and superlattices. 
   AFM topography of (b) undoped \ce{Sb2Te3-GeTe} and (c) Ti-doped \ce{Sb2Te3-GeTe} superlattice.
  }
  \label{fig_2}
\end{figure}

\begin{figure}[!htbp]
  \includegraphics[width=\textwidth]{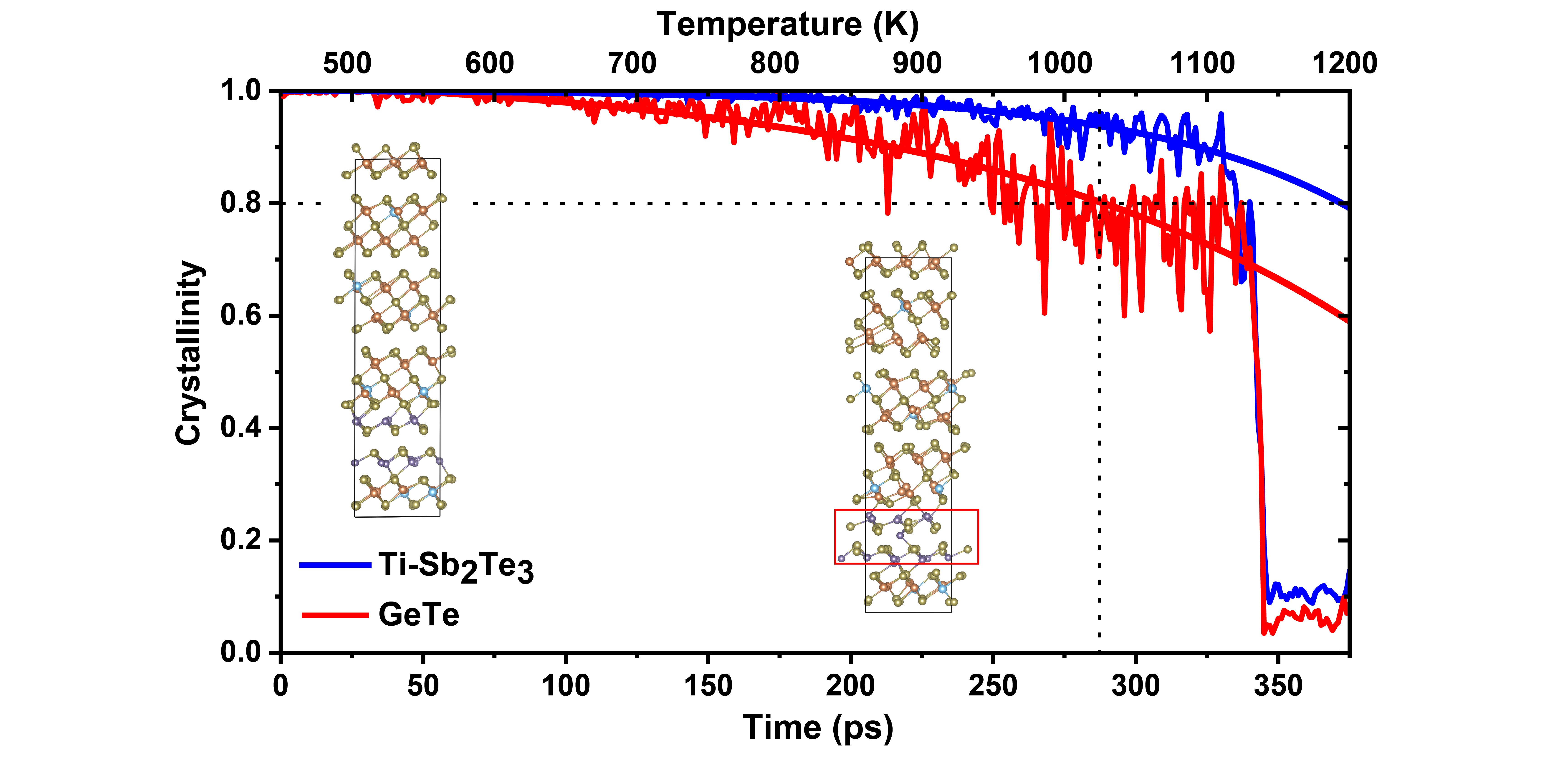}
  \caption{
   Simulated premelting behavior of Ti-doped \ce{Sb2Te3-GeTe} from 450 K to 1200 K.
   The inset shows the crystal structure at 450~K and 1135~K respectively.
   }
  \label{fig_3}
\end{figure}

\begin{figure}[!htbp]
  \includegraphics[width=\textwidth]{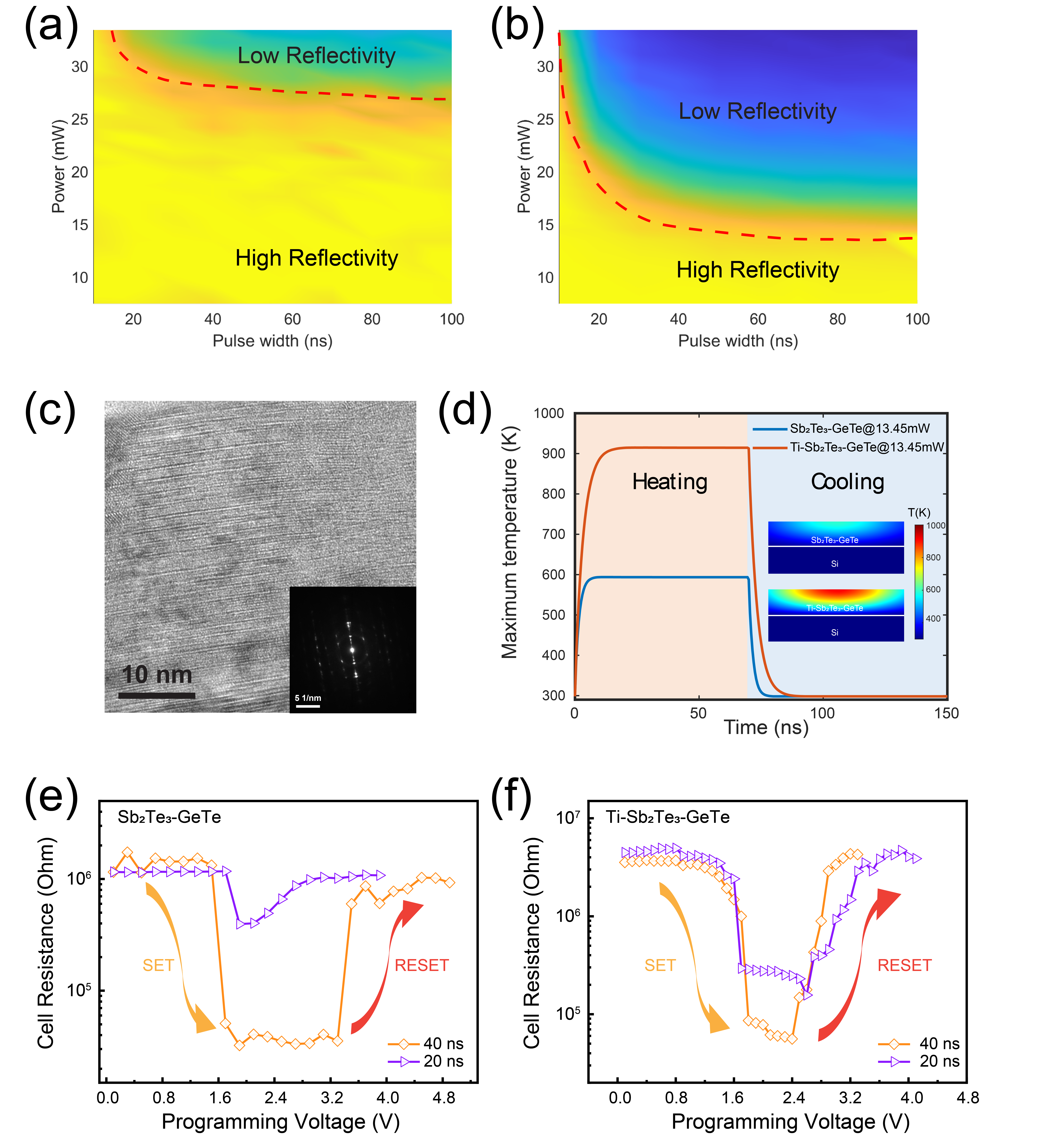}
  \caption{
  Laser RESET power-time-reflectivity measurement of (a) \ce{Sb2Te3-GeTe} and (b) Ti-doped \ce{Sb2Te3-GeTe} superlattice films.
  (c) cross-sectional TEM image of the area after laser switching pulse. The inset image is the selected area electron diffraction pattern.
  (d) Heat transport simulation of superlattice films during laser RESET process.
  Maximum temperature on the surface of superlattice films after a 70~ns laser pulse with different powers. The temperature distribution was shown in the embedded color map 
  The reversible resistance switching of R-V characteristics of the (e) \ce{Sb2Te3-GeTe} and (f) Ti-doped \ce{Sb2Te3-GeTe} superlattice films.
  }
  \label{fig_4}
\end{figure}

% Experimental section
\section{Methods}
\subsection{Superlattice growth and memory device fabrication}
The series of Ti-doped \ce{Sb2Te3} and Ti$_{3.6}$--(\ce{Sb2Te3})$_{96.4}$--GeTe superlattice films were deposited on Si (1~0~0) substrate by magnetron sputtering.
The background pressure of the main chamber was kept better than 6$\times$10$^{-5}$~Pa and the samples were prepared with an Ar atmosphere at a pressure of 0.5~Pa. 
The native silicon oxide layer on the Si substrate was removed by plasma etching at 25 W in an Ar atmosphere of 0.5 Pa for 60 min prior to the deposition. 
A 4-nm-thick amorphous \ce{Sb2Te3} was deposited at room temperature as seed layer and the further thin films were grown at 300$^{\circ}$C\citep{saito16, ning2020}. 
The films were cooled to room temperature in the vacuum chamber to avoid oxidation. 
The Ti-doped \ce{Sb2Te3} layers were grown by co-sputtering from \ce{Sb2Te3} and \ce{TiTe2} alloy targets (2'' diameter and 99.999\% pure from AJA International). 

The phase change random access memory devices were formed by alternately stacking 4-nm-thick \ce{Ti-Sb2Te3} and 1-nm-thick \ce{GeTe} layers on a mushroom-type device die.
This PCM device die was fabricated using standard CMOS processing.
The W bottom electrode was $\sim$210~nm in diameter.
The Ti-doped \ce{Sb2Te3-GeTe} superlattices were integrated into these prototype PCM devices by alternately stacking 4-nm-thick \ce{Ti-Sb2Te3} and 1-nm-thick \ce{GeTe} layers.
The total of 6 layer repeats were used to make a 30 nm thick superlattice film. 
Including the 4 nm seed layer, the total PCM thickness was 34~nm. 
A 20~nm thick TiN top electrode contact (TEC) was sputtered from a Ti sputtering target in a reactive Ar:N2 atmosphere at a pressure of 0.5 Pa. 
Finally, a 300~nm thick Al film was deposited on the TiN TEC to form a good electrical contact with the probing station. 
The Ti$_{3.6}$--(\ce{Sb2Te3})$_{96.4}$--GeTe superlattice phase change memory device was benchmarked against an undoped \ce{Sb2Te3-GeTe} superlattice. 
It was also grown on a 4 nm thick \ce{Sb2Te3} seed layer and was formed by alternately stacking 4~nm \ce{Sb2Te3} and 1~nm \ce{GeTe} layers using 6 repeats. 
With the exception of the Ti dopant, the devices were identical. 

\subsection{Characterization}
The crystal quality and orientation of films were measured by X-ray diffraction (XRD, Bruker D8 Discover) with Cu~K$_\alpha$($\lambda$=1.5418~\AA) radiation in a symmetric Bragg-Brentano geometry for $\theta$--$2\theta$ ranging from 5$^{\circ}$ to 60$^{\circ}$. 
The morphology and topography of the films were analyzed by field-emission scanning electron microscopy (SEM, JEOL-7600F) and atomic force microscopy (AFM, Asylum Research, MFP-3D).
The composition of the Ti-doped \ce{Sb2Te3} films was measured using energy dispersive spectroscopy (EDS) in the SEM. 
Raman spectra were obtained at room temperature by confocal Raman microscopy (WITec Alpha300R) with a 532~nm laser as the excitation source.
Thermal conductivities of the films were obtained by time-domain thermoreflectance (TDTR)~\citep{yamashita2021thermal,isosaki2017structure}.
%{harada2022crossover, wu2021interfacial}
The TDTR apparatus was equipped with an electrical delay control system, in which pulse emissions of two lasers were synchronized to an arbitrary function generator with 50~ns of delay range. 
The wavelength, repetition rate, pulse duration, average power, and spot radius of the pump laser were 1550~nm, 20~MHz, 0.5~ps, 20~mW, and 35~$\mu$m, respectively. 
For the probe laser, the wavelength, average power and spot radius were 775~nm, 1~mW, and 15~$\mu$m, respectively. 
The intensity of the pump laser was modulated by a rectangular wave of 200~kHz for lock-in detection. 
A 90-nm Al film was deposited on the top of samples as a transducer layer. 
The thermal conductivities of all samples were determined by fitting the measured thermoreflectance phase signals with those calculated from the delay time in the range from 0.2 to 50 ns.
The sample structure was measured using scanning transmission electron microscopy (S/TEM, JEOL/FEI Titan) with an acceleration voltage of 200~kV, as shown in Fig.~\ref{fig_1}(d) and \ref{fig_4}(c). 
Focused Ion Beam (FIB, FEI Helios Nanolab 450S) milling was necessary to prepare the lamella for cross-sectional TEM image. %G4 CX DualBeam\textsuperscript{TM}
The optical switching was carried out using our home-made static tester, which consists of a high-power 658~nm pump laser and a relatively low-power 635~nm probe laser, which can measure the switching power and time~\citep{behera2017laser}. 
The system can simultaneously capture the reflection of the probe laser from the sample whilst the pump laser pulses heat into the sample.
The focused laser spot had a beam size of 0.6~$\mu$m ($1/{e^2}$ intensity) on the sample. %~\citep{suzaki1975measurement}. 
%An array of RESET and SET marks under different laser pulse widths and incident powers was made on the surface of samples. 
%The reflected signal from the probe laser was collected before and after the pump pulses.
The electrical switching performance of the devices was measured using a homemade system with a sourcemeter (Keithley 2600) and pulse generator (Tektronix AWG5002B). 
The cell resistance of the devices was recorded after each programming pulse at a constant voltage of 0.1~V in order to minimize the read disturbance in the device.

\subsection{Modelling \& Simulaitons}
\subsubsection{DFT Molecular Dynamic Calculations }
DF/MD calculations were performed using the Vienna $ab initio$ Simulation Package (VASP)~\citep{Kresse1993} with projected augmented wave (PAW) pseudopotentials~\citep{Kresse1999} and Perdew-Burke-Ernzerhof generalised gradient approximation (PBE--GGA) exchange-correlation functional~\citep{Perdew1996}. 
The spin-orbit interaction was neglected and the vdW interaction correction method of Grimme was included~\citep{grimme2010consistent}.
The unit cell of \ce{Sb2Te3} with three quintuple blocks was used to explore the potential Ti sites when doping. 
The free energy and formation energy calculations were carried out at the $\Gamma$ point in the Brillouin zone (k=0). 
The plane-wave cutoff energy was 300 eV. 
The chemical bonding was studied by crystal orbital Hamiltonian population (COHP) analysis with LOBSTER program~\citep{dronskowski1993crystal}.

The MD premelting study used a [Ti-\ce{Sb2Te3}]$_{\textmd{4nm}}$--[\ce{GeTe}]$_{\textmd{1nm}}$ supercell $(3\times3\times1)$ consisting of 9 Ti, 18 Ge, 63 Sb and 126 Te atoms,, as shown in Fig.~\ref{fig_1}(a).
The configuration of the Ti atoms, which substituted Sb atoms, were searched using \texttt{pymatgen} package~\citep{ong2013python}.
The geometry relaxation of the supercell was carried out at 0~K after which NVT canonical ensemble was used to update the atomic positions every 3 fs whilst the relaxed structure was heated at 2~K/ps from 450 K  to 1200 K, which is above the structure's melting point, as seen in Fig. \ref{fig_3}.
The initial velocities at 450~K  were random, and scaled according to the temperature. 
The temperature of the model during the temperature ramp was controlled by velocity rescaling. 
The energy of the ensemble was computed at the $\Gamma$ point in the Brillouin zone (k=0). 
The plane-wave cut-off energy for the the molecular dynamics simulations was 240 eV. 

The \ce{GeTe} premelting behaviour was studied by calculating the Bond Orientation Parameter (BOP) of the \ce{Ti-Sb2Te3} and GeTe sections of the simulation supercell, which is commonly used to identify the different local atomic crystal structures and to study structural phase transitions~\citep{steinhardt1983bond, kalikka2012nucleus}. 
The local structure around atom $i$ is defined by a set of spherical harmonics, where $Y_{lm}(\widehat{\textbf{r}}_{ij})$, $\widehat{\textbf{r}}_{ij}$ is the vector between atom $i$ and one of its neighbours $j$.
$N_b$ is the total number of neighbours around the $i^{th}$ atom. 
The global bond order parameter, $Q_l$, can be then calculated by averaging according to:  
\begin{equation}
	Q_{l} \equiv {\sqrt{ {\frac{4\pi}{2l+1} \sum_{m=-1}^{l}
	 {\left| 
	 {\frac{1}{N_b}
	 {\sum_{j=1}^{N(i)}
	 Y_{lm}(\widehat{\textbf{r}}_{ij})}}\right| }^2}}}
	\label{eq_bop}
\end{equation}
where $l=4$ for a local atomic cubic structure. 
The layer crystallinity was calculated as an average over all the atoms occupying the space initially occupied by each layer, and over 1~ps (2~K) windows. 

\subsubsection{Laser-heat-induced superlattice switching simulations}
Heat generated by laser pulses in the superlattice was used to induce interfacial phase transitions, see Fig. \ref{fig_4}. 
We modelled the transient heating effect using a finite-difference time-domain approach to solve the unsteady heat conduction equation, as expressed in Eq.~\ref{eq_Temp},
\begin{equation}
	\rho c_p \frac{\partial{T(x,y,z,t)}}{\partial t}= \triangledown\cdot\kappa\triangledown T(x,y,z,t)+Q(x,y,z,t)
	\label{eq_Temp}
\end{equation}
where, $\rho$ is the mass density, $c$ is the specific heat capacity, $T(x, y, z, t)$ is the temperature of a position of $(x, y, z)$ at a certain moment $t$, $\kappa$ is the thermal conductivity, $Q(x, y, z, t)$ is the Joule heat brought by the laser pulse with a Gaussian beam profile, which can be given as Eq.~\ref{eq_energy}, 
\begin{equation}
	Q(x,y,z,t)=e^{-\alpha z}\frac{2P_{in}}{\pi w^2}(1-R)\alpha e^{-2\frac{x^2+y^2}{w^2}}f(t)
	\label{eq_energy}
\end{equation}
where, $P_{in}$ is the laser power, $w$ is the $1/e^2$ Gaussian beam radius, $\alpha$ is the absorption coefficient, $R$ is the reflectivity, and $f(t)$ is the temporal waveform. 
Here, the heat capacity ($\rho c_p$) of \ce{Ge2Sb2Te5}($\sim$1.25$\times$10$^6$~J$\cdot$m$^{-3}\cdot$K$^{-1}$) was used~\citep{reifenberg2006multiphysics}.
$\alpha$ of \ce{Sb2Te3-GeTe} ($\sim$8.155$\times$10$^7~$m$^{-1}$) and Ti-doped \ce{Sb2Te3-GeTe} ($\sim$5.676$\times$10$^7$~m$^{-1}$) were calculated from measured extinction coefficient (Fig.~\ref{fig_s_nk}).
$R$ of \ce{Sb2Te3-GeTe} (0.63) and  Ti-doped \ce{Sb2Te3-GeTe} (0.50) superlattices were measured with a 660-nm laser, which are consistent with the simulated reflection from measured refractive index and extinction coefficient, as shown in Fig.~\ref{fig_s_nk}.
We measured $\kappa$ of \ce{Sb2Te3-GeTe} ($\sim$0.49~W$\cdot$m$^{-1}\cdot$K$^{-1}$) and Ti-doped \ce{Sb2Te3-GeTe} ($\sim$0.24~W$\cdot$m$^{-1}\cdot$K$^{-1}$) using TDTR system (Fig.~\ref{fig_2}(a)).
A 70ns laser pulse was used to simulate the temperature of the film as a function of time for different incident laser powers.

\backmatter

\bmhead{Supplementary Information}
Whilst writing this article we became aware of a similar set of results from the group of Mann-ho Cho at Yonsei University, Korea. 
See REF XXX.
The fact that two different research groups independently show similar increases in material switching efficiency and stability using different techniques lends further credibility to the results presented herein.
\\

\noindent
This article has an accompanying supplement. 

\bmhead{Acknowledgments}

The SUTD-based research was funded by two different grants: (1) Singapore Ministry of Education (MoE) award \# 2017-T2-1-161, and (2) Office of Naval Research Global award \# N62909-19-1-2005.
Ms Jing Ning is grateful for her MoE PhD scholarship.

\bibliography{Ti_GST_SL_refer}% common bib file

\clearpage

\clearpage

\setcounter{figure}{0} 
\renewcommand{\thefigure}{S\arabic{figure}}
\section{Supplementary Information }

\begin{figure}[!htbp]
  \includegraphics[width=\textwidth]{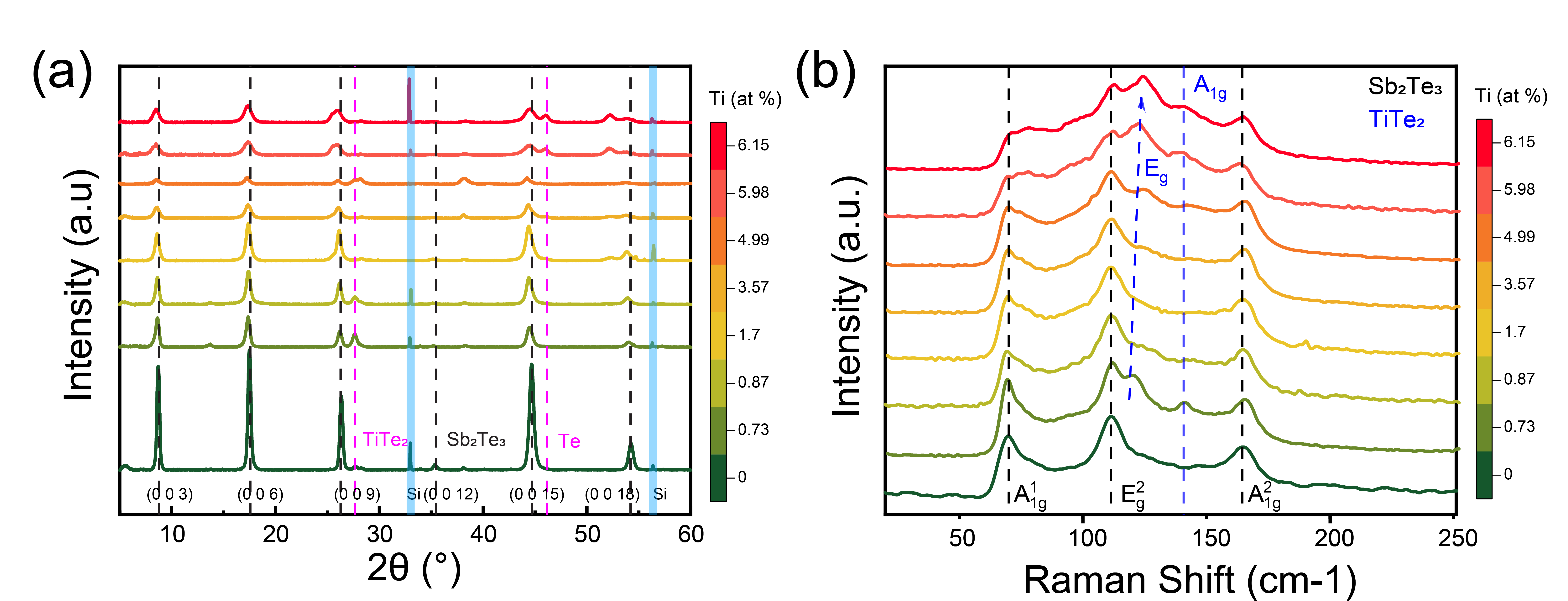}
  \caption{
  (a) XRD patterns and (b) Raman spectra of pure \ce{Sb2Te3} and doped \ce{Sb2Te3} films with different concentration of Ti.
  The concentration of Ti was shown in the color range from green to red.
  }
  \label{fig_S_XRD_Raman}
\end{figure}

\begin{figure}
  \includegraphics[width=\textwidth]{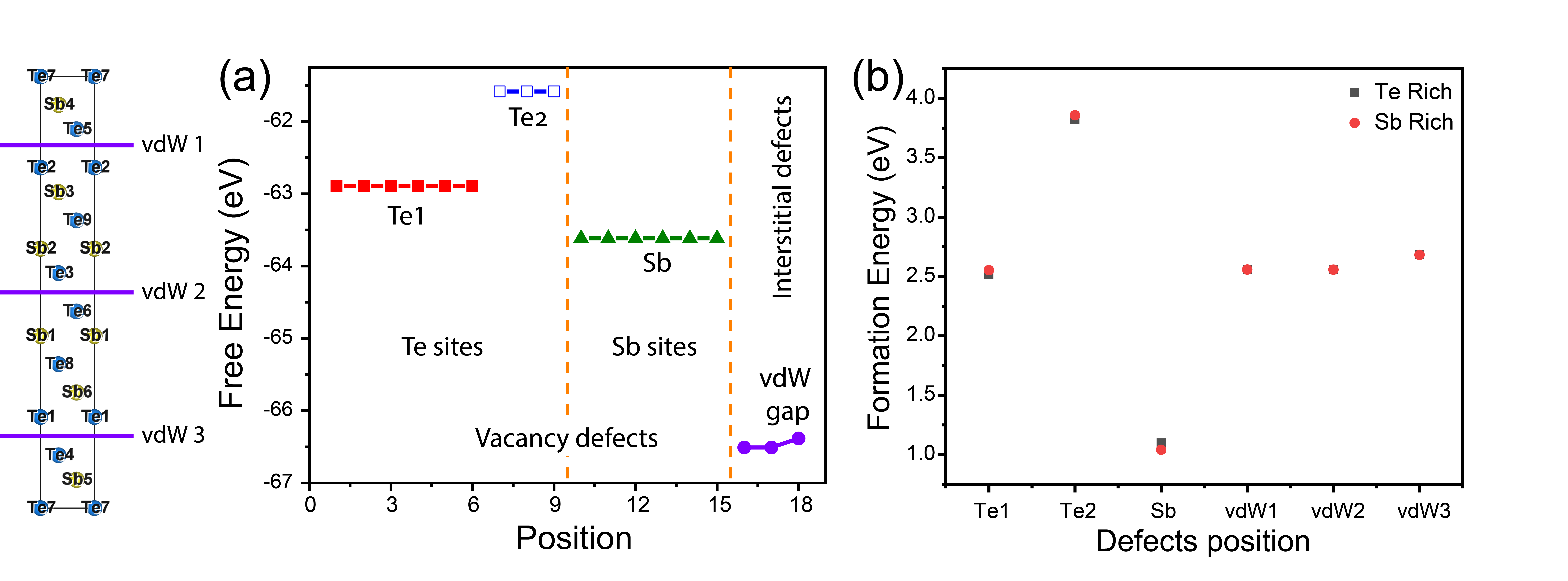}
  \caption{
  (a) Free energy and
  (b) Formation energy calculation of structures with different Ti defects.
  Defects sites was labelled in the \ce{Sb2Te3} structure. 
  }
  \label{fig_S_energy}
\end{figure}

\begin{figure}
  \includegraphics[width=\textwidth]{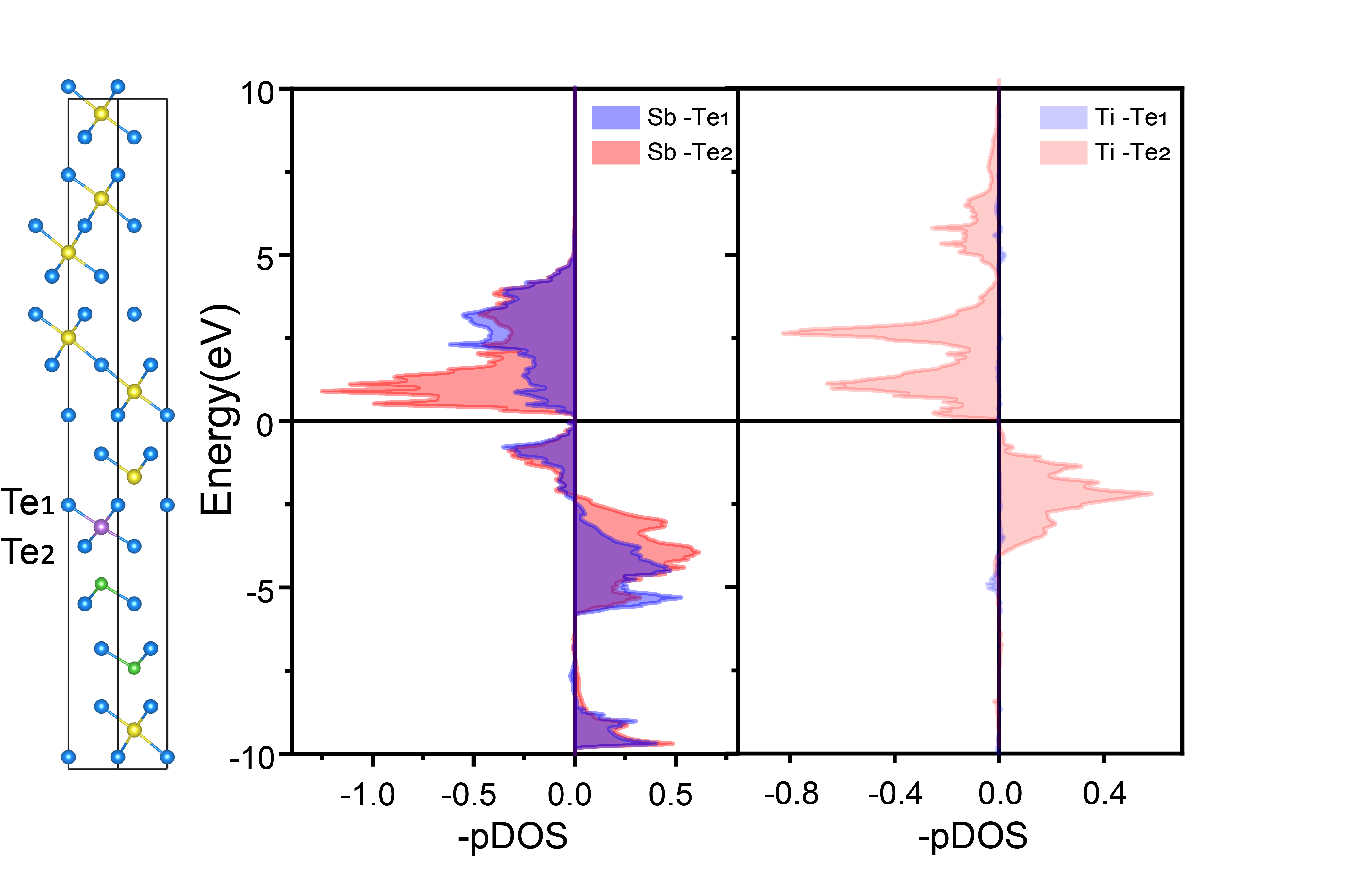}
  \caption{
   Bonding states comparison of Sb or Ti and its neighbors in superlattice structure.
   }
  \label{fig_S_COHP}
\end{figure}

\begin{figure}
  \includegraphics[width=\textwidth]{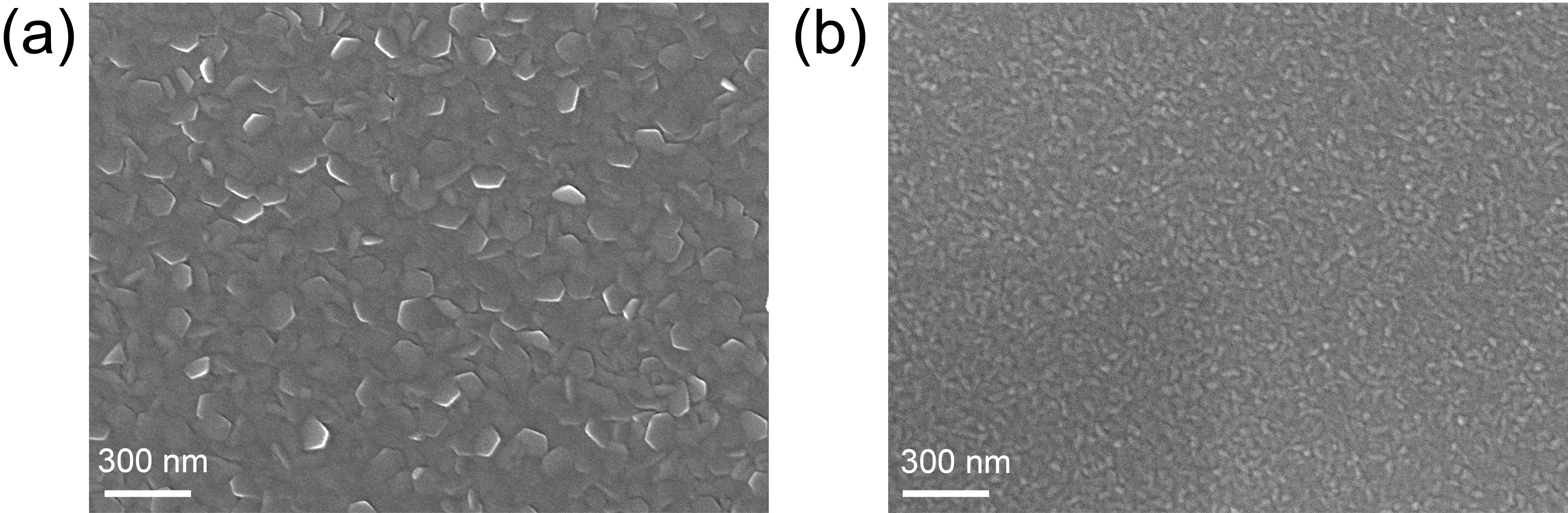}
  \caption{
  Morphology of (a) \ce{Sb2Te3-GeTe} and (b) Ti-doped \ce{Sb2Te3-GeTe} superlattice films.
  }
  \label{fig_S_SEM}
\end{figure}

\begin{figure}
  \includegraphics[width=\textwidth]{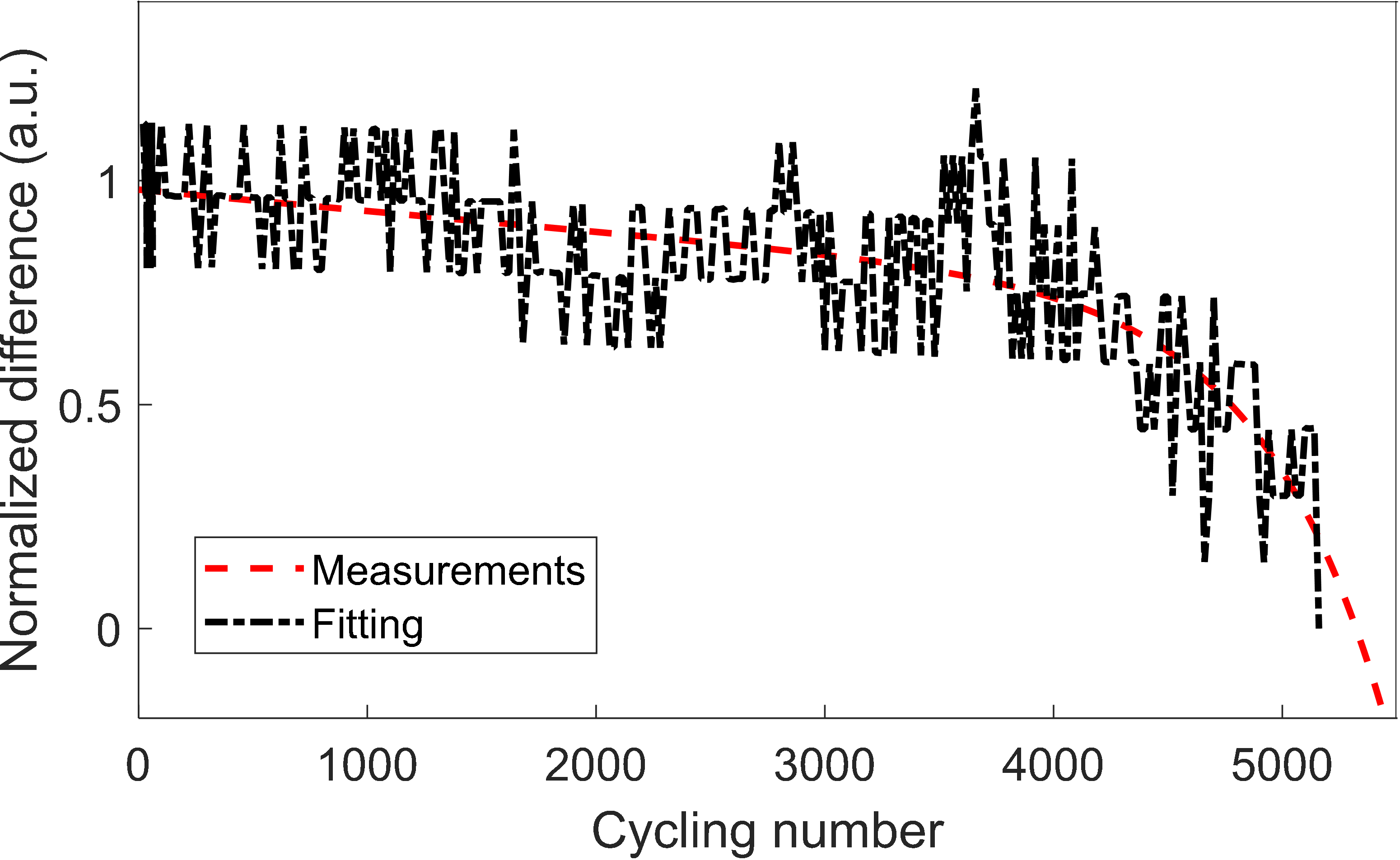}
  \caption{
  The endurance-cycling measurements for the Ti-doped \ce{Sb2Te3-GeTe} superlattice structure using 50ns, 13.45~mW and 300~ns, 9.42~mW as RESET and SET pulses, respectively. 
  }
  \label{fig_S_cycle}
\end{figure}

\begin{figure}
  \includegraphics[width=\textwidth]{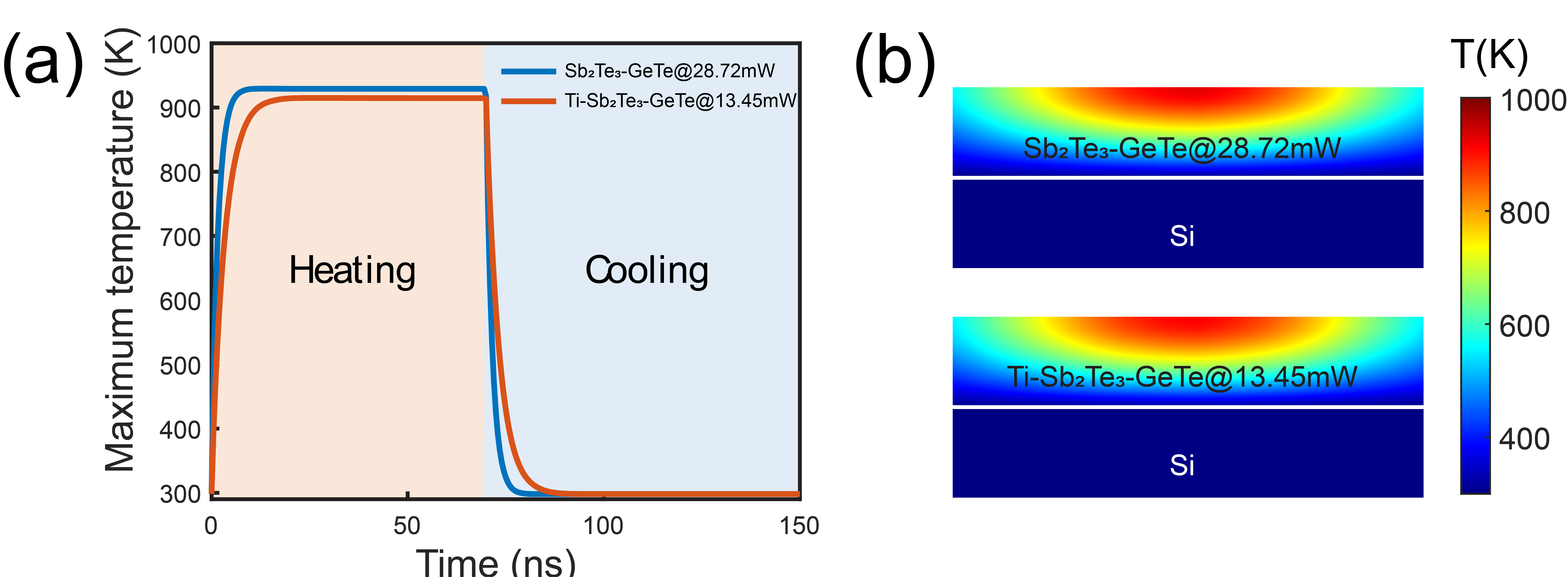}
  \caption{
  Heat transfer simulation of \ce{Sb2Te3-GeTe} and Ti-doped \ce{Sb2Te3-GeTe} in RESET operation.
  }
  \label{fig_S_heat}
\end{figure}

\begin{figure}
  \includegraphics[width=\textwidth]{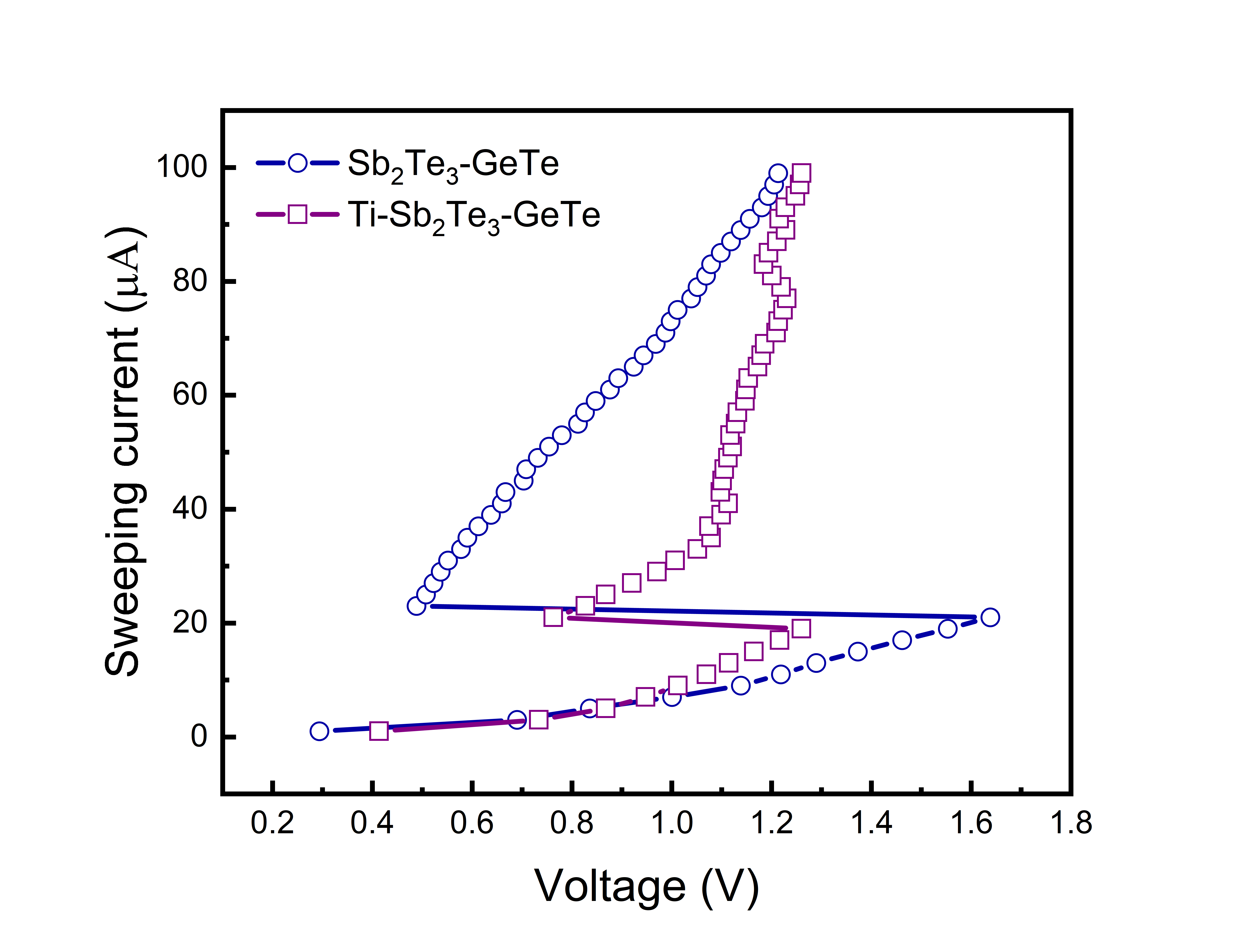}
  \caption{
  The threshold switching characteristics of the undoped and Ti-doped \ce{Sb2Te3-GeTe} superlattice iPCM devices.
  }
  \label{fig_S_IV_curve}
\end{figure}

\begin{figure}
  \includegraphics[width=\textwidth]{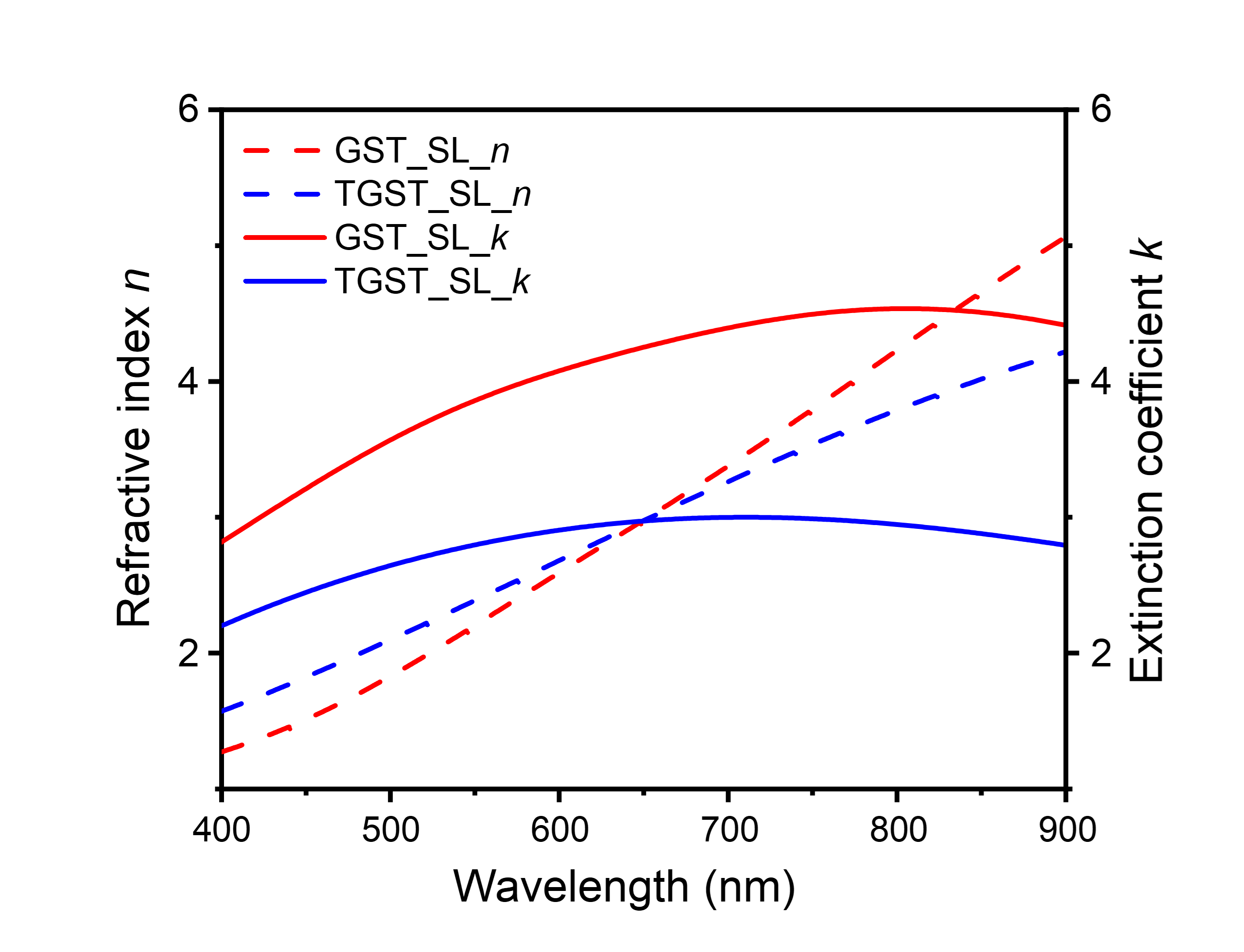}
  \caption{
  Optical properties of two superlattice films. 
  Reflective index, n, and extinction coefficient is displayed as dash-line and line,  respectively.
  }
  \label{fig_s_nk}
\end{figure}

\end{document}